\documentclass[prb,twocolumn,groupaddress]{revtex4}

\usepackage{amsmath} 
\usepackage{amssymb} 
 \usepackage{amsfonts}

\usepackage{graphicx} 

\usepackage{array}
\usepackage{multirow}
\begin{document}

\title{Sorption of proteins to charged microgels: characterizing binding isotherms and driving forces} 

\author{ Cemil Yigit, Nicole Welsch, Matthias Ballauff, and Joachim Dzubiella$^*$}
\affiliation{Soft Matter and Functional Materials, Helmholtz-Zentrum Berlin, Hahn-Meitner Platz 1, 14109 Berlin, Germany}
\affiliation{Department of Physics, Humboldt-University Berlin, Newtonstr.~15, 12489 Berlin, Germany}

\thanks{To whom correspondence should be addressed. E-mail: joachim.dzubiella@helmholtz-berlin.de}

\begin{abstract}

We present a set of Langmuir binding models in which electrostatic cooperativity effects to protein sorption is 
incorporated in the spirit of Guoy-Chapman-Stern models, where the global substrate (microgel) charge state
is modified by bound reactants (charged proteins). Application of this approach to 
lysozyme sorption to oppositely charged core-shell microgels allows us to extract the intrinsic, 
binding affinity of the protein to the gel, which is salt-concentration independent and mostly hydrophobic in nature. 
The total binding affinity is found to be mainly electrostatic in nature, changes many orders of magnitude 
during the sorption process, and is significantly influenced by osmotic deswelling effects. The intrinsic binding
affinity is determined to be about 7 $k_BT$ for our system. We additionally show that Langmuir 
binding models and those based on excluded-volume interactions are formally equivalent for low to moderate 
protein packing, if the nature of the bound state is consistently defined. Having appreciated this, a more 
quantitative interpretation of binding isotherms in terms of separate physical interactions is possible in future for 
a wide variety of experimental approaches. 

\end{abstract}

\maketitle

\section{Introduction}

Functionalized colloids and nanoparticles play an increasingly 
prominent role in the development of biomaterials and imminent 
biotechnological applications, for example, drug delivery, enzyme biocatalysis, or control of gene expression.~\cite{peppas,caruso,alexander, bajpai,levental, linse:pnas, linse:nanoletters,haag,dawson1}  In particular, multiresponsive hydrogels are of great interest due their
biocompatibility, resemblance to biological tissue, and tunable viscoelastic 
properties.~\cite{peppas,alexander,bajpai,levental,eichenbaum,eichenbaum2,sassi,khoury,lyon,dawson2} 
Dispersed in water these colloidal microgels create an enormous surface that may be taken as a model for soft biological interfaces. Hence, protein storage, activity, 
and uptake properties may be changed by physiological stimuli, such as pH, 
salt concentration, and temperature. Moreover, colloidal hydrogels qualify for a number of applications, e.g., 
as drug carrier devices.~\cite{blackburn,smith,ghugare}   However, the detailed control of
functionality  requires a  quantitative understanding  
of the underlying physical interactions between
microgels and  biomolecules in the  aqueous environment.

Recent studies have demonstrated that protein sorption 
to nanoparticles is mostly driven by global, nonspecific electrostatic interactions 
and more local, probably hydrophobic  interactions.~\cite{linse:pnas,linse:nanoletters,sassi, kabanov, rotello, kissel,kissel2, nicole, hansson:2007,hansson,norde}  The balance between those two is highly system-specific  and can be manipulated  
by chemical functionalization or copolymerization. Charged microgels, for 
instance, can be used to favor or disfavor the sorption of net-charged 
proteins, while their osmotic swelling and storage volume can be tuned
by pH, salt and charge density~\cite{eichenbaum,eichenbaum2, cruz, hansson} essentially via the
Donnan equilibrium.~\cite{flory} However, during the uptake of the charged, polyionic proteins, 
swelling and Donnan equilibria are typically changing in an 
interconnected fashion.~\cite{kabanov, hansson:2007,hansson,norde} These highly {\it cooperative} effects 
render the interpretation of binding isotherms, and thus the separation and quantification of 
global electrostatic and local hydrophobic contributions to binding, a difficult task. Moreover, binding affinities 
in these systems depend on protein load which presents an additional complication when modeling 
the adsorption isotherm.

Up to now, the modeling of protein uptake
 has been done often by using the standard Langmuir isotherm,\cite{langmuir,langmuir2} 
in particular when evaluating protein adsorption as measured by isothermal titration calorimetry (ITC).~\cite{itc,linse:pnas,linse:nanoletters, rotello,kissel2, nicole}
In the standard Langmuir approach, however, protein association with single, independent 
binding sites is assumed which neglects electrostatic cooperativity effects and volume
changes during sorption. Additionally the term 'binding' of proteins to soft 
polymeric layers and hydrogels is somewhat ill defined, as the system
may remain in a fluid-like state where proteins are still mobile on 
average, albeit slower than in bulk.~\cite{khoury,norde:mobility} 
Consequently the stochiometry and binding affinities to 'sites' in the 
gel obtained from Langmuir fitting are not  so easy to interpret.
 
Alternatively, the hydrogel matrix may be viewed as a 
homogeneously charged background to the mobile proteins where saturation
of sorption may set in due to excluded-volume (EV) packing. In recent binding models based 
on such a view,~\cite{biesheuvel:pre,vos:langmuir,cellmodel:wittemann,hansson}
it is typically assumed that  proteins can be treated as simple charged hard spheres, i.e., hard polyions. 
The electrostatic problem can then be  tackled by approximative Poisson-Boltzman (PB) 'cell' or 'box' models, ~\cite{wennerstrom1,wennerstrom2,cellmodel:pincus, cellmodel:denton, biesheuvel:pre,vos:langmuir,cellmodel:wittemann, barbosa,deserno,allen:warren,hansson:jcis} where electrostatic cooperativity and osmotic ion effects 
to deswelling can be included. So far it has not been attempted, however, to separately treat more specific effects 
in EV-based binding models such as hydrophobic interactions or restraints of the configurational 
protein degrees of freedom in the bound state. Also the relation between Langmuir and EV models, 
if any, is unclear. However, modeling protein adsorption by a meaningful and physically sound isotherm is the prerequisite for a quantitative understanding of the driving forces.

Here we present an in-depth discussion of the isotherms suitable for modeling protein sorption into microgels and soft polymeric layers in general. This discussion will then provide the basis for a detailed investigation of the driving forces of protein binding. We demonstrate that electrostatic  cooperativity and the effects of microgel volume changes 
can be introduced into standard Langmuir models following the spirit of Guoy-Chapman-Stern 
theory for binding of charged molecules to  
charged surfaces.~\cite{haydon,seelig,mclaughlin} We test the performance of the binding 
models by fitting to  binding isotherms obtained previously from ITC
of chicken egg white lysozyme sorption onto a negatively charged coreÐshell 
microgel.\cite{nicole} The core-shell particles consist of a polystyrene core onto 
which a charged poly (N-isopropylacrylamide-co-acrylic acid) (NiPAm) network is attached. 
Previous characterization of this system shows that it is an ideal model system
as it reaches full equilibrium with high binding affinities, and the globular lysozyme 
maintain its folded state in the microgel with even enhanced activity.~\cite{nicole} 
We then demonstrate that we can separate out the global electrostatic contribution 
leading to consistent values for the salt concentration independent binding affinity. 
Finally we show that EV-based binding models are formally equivalent to the Langmuir
approach in the low-packing regime if the bound state is consistently defined.
Consequences to the interpretation of Langmuir models are discussed.

\section{Experiments: Materials and Methods}

\subsection{Materials}

In this study the same batch of microgel dispersion was used as in previous work.~\cite{nicole} 
In brief, the polystyrene core was synthesized by emulsion polymerization in the first step. After purification of the core particles, the microgel shell, containing 5 mol-\% N,NÕ-methylenebisacrylamide (BIS) crosslinkers and 10 mol-\% acrylic acid with respect to the amount of NiPAm, was polymerized on the polystyrene core by seed polymerization. After purification the particles were transferred into buffer solution by ultrafiltration against 10 mM MOPS buffer at pH 7.2. In this preparation state the gel is swollen and strongly hydrated with more  than 90\% volume fraction of water.  
Dynamic Light Scattering (DLS, Malvern Instruments) was applied to determine the hydrodynamic radius of the polystyrene 
core to $R_{\rm core}=62.2 \pm0.7$~nm and the radius of the total core-shell microgel to $R\simeq 129$ to $172$~nm 
depending on the solution condition.\cite{nicole} 

The ITC experiments were performed using a VP-ITC instrument (Microcal) as described previously.~\cite{nicole} Briefly, a 
total of 300 ml of lysozyme solution (0.695 mM) was titrated into the sample cell filled with 1.4~ml of buffer-matched microgel dispersion at a $c_m=8.42\cdot10^{-7}$~mM concentration. The pH-value was held constant at pH=7.2. The experiments were performed at 298 and 303 K. 
The following three buffer systems were used: (i) 10 mM MOPS, 2 mM NaN$_3$ (7 mM ionic strength); (ii) as (i) with additional 10 mM NaCl (17 mM ionic strength); (iii) as (i) with additional 25 mM NaCl (32 mM ionic strength). The incremental heat changes $\Delta Q$  were measured during the course of the titration experiment where the protein solution is stepwise injected into the microgel dispersion. The integrated heat change after each injection was corrected by the heat of dilution.
 
\section{Theory}

\subsection{Basic model}

\begin{figure}[h]

\centerline{\hspace{0.3cm}\includegraphics[width=10.0cm,angle=0]{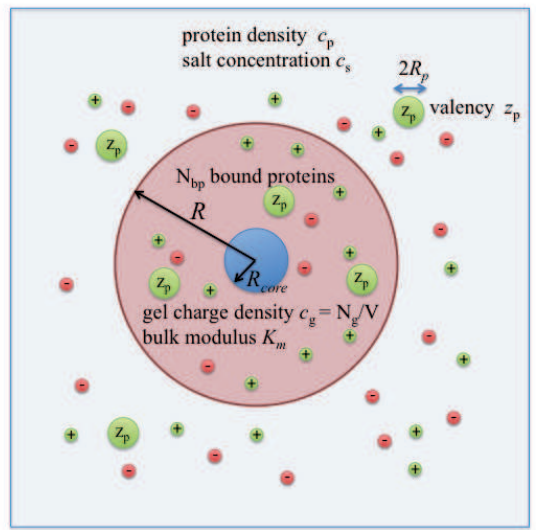}}
\caption{Illustrative sketch of the model for one core-shell microgel in a protein dispersion. The microgel is 
represented by a sphere with radius $R$ between 129 and 172~nm 
including a core with radius $R_{\rm core}=62.2$~nm.  The gel with volume 
$V$ has a mean charge density  $c_g=N_g/V$.  Protein and monovalent salt concentrations in the bulk region  are $c_p$ and $c_s$, respectively. The proteins have radius $R_p$ and valency $z_p$. $N_{bp}$ is the number of bound proteins.}
\label{lab_snapbulk}
\end{figure}

In our model one core-shell hydrogel particle is 
modeled as a perfect sphere with radius $R$ having a core  with radius $R_{\rm core}$ as depicted in 
Fig.~1.  The net gel volume is therefore $V=4\pi(R^3-R_{\rm core}^3)/3.0$. Potentiometric measurements 
show that a total number of $N_g = 4.9\cdot10^{5}$ charged monomers  are present in the gel.  
The network monomers are assumed to be homogeneously distributed in the gel as justified by small
angle scattering.~\cite{profile} The mean gel 
charge density is $z_gc_g = z_gN_g/V$, where 
$z_g$ is the monomer charge valency.  Since we work at a pH $\simeq 7.2$, much larger than the pK$_a$-value 
of $\simeq 4.6$ of polyacrylic acid, and deal with weakly charged gels (with a charge fraction $\simeq 1/10$), we can safely assume that charge regulation plays a minor role and $z_g=-1$ in the 
following.  The salt is monovalent with a bulk concentration $c_s = c_+ = c_-$, where $c_+$ and $c_-$ are the 
concentrations of cations and anions, respectively.  The
Bjerrum length is $l_B = e^2/(4\pi\epsilon\epsilon_0 k_BT)$, where $e$ is the elementary charge, $\epsilon_0$ and
$\epsilon(T)$ the vacuum and bulk permittivities, respectively, and $k_BT=\beta^{-1}$ the thermal energy. In our systems
$l_B\simeq 0.7$~nm, while the electrostatic Debye-H{\"u}ckel screening length $\kappa_b^{-1} = (8\pi l_B c_s)^{-1/2}$ 
in bulk is between $\simeq$~3.7~nm (7 mM salt), $\simeq$~2.4~nm (17 mM salt),  and $\simeq$~1.7 nm (32 mM salt).  

The lysozyme proteins are modeled by monopolar, homogeneously  charged spheres with diameter $\sigma_p=2R_p$, assumed to have a valency  $z_p=+7$ (i.e., a net charge of +7$e$) estimated
from titration.~\cite{kuehner} Hence, multipolar contributions, e.g., dipolar or quadropolar or charged-patch 
interactions to sorption are neglected.
The proteins are found in bulk solution with density $c_p$. No dispersion or other attractive nonelectrostatic interactions between the proteins are considered, and we assume a homogeneous distribution of proteins in the gel region.~\cite{nicole}
Aggregation of proteins in the gel is unlikely because at full load the system is still
below the solubility threshold of a bulk system at comparable pH and electrolyte 
concentration.~\cite{ducruix} The number of bound protein inside the gel is denoted by $N_{bp}$ giving rise to 
an internal protein packing fraction $\eta = N_{bp}\pi\sigma_p^3/{6V}$. Finally, the gel and the aqueous buffer are 
modeled as a  continuum background  with elastic, dielectric,  and osmotic properties as detailed in the 
following  sections. The introduced quantities and variables are summarized in Tab.1.

\begin{table}[h]
\begin{center}
\begin{tabular}{|c|c|}
\hline
variable & meaning \\
\hline
$R$ & gel radius\\
$V$ & gel volume\\
$c_g$ & charged monomer concentration\\
$c_s$ & bulk salt concentration\\
$\kappa_g$ & inverse Debye length in gel\\ 
$\kappa_b$ & inverse Debye length in bulk \\ 
$c_p$ & bulk protein concentration\\
$c_p^{\rm tot}$ & total protein concentration\\
$x$ & molar ratio $x =c_p^{\rm tot}/c_m$\\ 
$N_{bp}$ & number of proteins in the gel\\
$N$ & number of Langmuir binding sites\\
$\Theta$ & protein load $\Theta = N_{bp}/N$\\
$\sigma_p$ &  effective protein diameter \\
$R_p$ &  effective protein radius\\
$\eta$ & protein packing fraction in gel\\
$\Delta\phi$ & Donnan potential\\
$K$ & binding constant \\
$K_m$ & gel modulus\\
\hline 
constants & meaning \\
\hline
$R_{\rm core}=62.2$~nm & core radius\\
$N_g=4.9\cdot10^5$ & number of charged monomers\\
$z_g=-1$ & monomer charge valency\\
$z_p= 7 $  & protein charge valency\\
$c_m=8.42\cdot10^{-10}$~M & microgel concentration\\
$\l_B\simeq0.7$~nm & Bjerrum length\\
$v_0$=l/mol & standard volume\\
\hline
\end{tabular}
\end{center}
\caption{Frequently used variables and constants.}
\label{table1}
\end{table}

\subsection{Electrostatic contributions to the interactions of the proteins with the charged gel}

\subsubsection{Donnan equilibrium}

Homogenization of the charged gel by salt ions leads to a electrostatic potential difference
between bulk and gel, the Donnan potential.
The latter can be  derived by assuming electroneutrality in the gel for monovalent, ideal ions leading
to~\cite{flory} 
 \begin{eqnarray}
\Delta \tilde \phi \equiv e\beta\Delta\phi = \ln\left[y+\sqrt{y^2+1}\right], 
\label{donnan}
\end{eqnarray}
where $y = z_g c_g/(2c_s)$ is the charge ratio between gel and bulk charge densities and the tilde
in $\Delta\tilde\phi$ denotes the dimensionless potential scaled by $e\beta$.

We estimate a mean separation of about $2.2$~nm between two charged monomers on the same polymer within the gel,  considerably larger than the Bjerrum length  $l_B\simeq0.7$~nm.  Thus, charge regulation effects by counterion 
(Manning) condensation or inhomogeneity effects can be safely neglected.~\cite{manning:donnan, buschmann, khokhlov, dobrynin}  For our microgel we further estimate the charged monomer concentration to be 
$c_g = N_g/V\simeq 40 - 100$~mM, depending on the swelling state. 
Given bulk salt concentrations between 7 and 32 mM,  thus $|\Delta\tilde\phi|$ is 
on the order of unity ($\simeq$~25~mV) and significant for the interaction with charged macromolecules.

The Donnan equilibrium leads to an osmotic pressure difference $\pi_{\rm ion}$ between gel and bulk ions and, in 
the ideal gas limit, is given by the difference of internal and external ionic concentrations~\cite{flory}
\begin{eqnarray}
\label{ionpressure}
\beta \pi_{\rm ion} &=& c_se^{-\Delta\tilde\phi}+c_s e^{\Delta\tilde\phi}-2c_s\\  \nonumber
&=& 2c_s\left[\sqrt{y^2+1}-1\right] = 2c_s[\cosh\Delta\tilde\phi-1].
\end{eqnarray}
For the salts and concentrations considered in this work corrections due to 
nonideal activity are small~\cite{robinson} and can be safely neglected.

\subsubsection{Electrostatic protein-gel interaction}

The transfer of a charged protein from the bulk solution with salt concentration $c_s$ 
into the charged gel with monomer charge concentration $c_g$ is accompanied by a (Gibbs) transfer free 
energy change per particle (or chemical potential). If we assume the gel monomers to 
be mobile, this can be modeled by the transfer of a spherical polyion with radius $R_p$ and 
valency $z_p$ from one bath at salt concentration $c_s$ and zero potential to another at mean potential
$\Delta \phi$ and salt concentration $c_g$. The difference in solvation free energies on a Debye-H\"uckel level is then~
\begin{eqnarray}
\beta \Delta G_{el} = z_p\Delta\tilde\phi(y)& -{\displaystyle \frac{z_p^2l_B}{2R_p}\left(\frac{\kappa_g R_p}{1+\kappa_g R_p}-\frac{\kappa_b R_p}{1+\kappa_b R_p}\right)} 
\label{transfer}
\end{eqnarray}
where we defined the inverse screening length $\kappa_g = \sqrt{8\pi\l_B c_g}$ in the gel, characterizing 
approximately the screening by mobile polyelectrolyte charges and their neutralizing counterions (see Appendix A). 
The  Donnan potential (\ref{donnan}) must now be corrected for the change of total net charge by the 
charge of bound protein via
\begin{eqnarray}
y = (z_g c_g + z_pN_{bp}/V)/(2c_s).
\label{ratio}
\end{eqnarray}
The first, purely entropic term in eq.~(\ref{transfer}) simply expresses 
the electrostatic transfer energy of a charge $z_p$ 
from bulk to a region at potential $\Delta\tilde\phi(y)$. The second term
reflects the difference in the Born solvation free energies in a salty medium.~\cite{mcquarrie} 
The first term is attractive if microgel and protein have opposite
net charge, otherwise repulsive, while the second one is attractive if $c_g > c_s$, otherwise repulsive. 
Since we deal with a weakly charged gel additional entropic effects from the release of condensed
counterions can be neglected in our system.~\cite{henzler}

In Appendix~A a more rigorous derivation of eq.~(\ref{transfer}) is presented in the
framework of a linearized PB cell model~\cite{wennerstrom1,wennerstrom2,cellmodel:pincus, cellmodel:denton, biesheuvel:pre,vos:langmuir,cellmodel:wittemann, barbosa,deserno,allen:warren,hansson:jcis} corrected for 
particle-particle interactions.  While eq.~(\ref{transfer}) is valid strictly only in the limits of small 
salt concentrations and small protein charges, its 
quantitative validity  seems to cover a wide range of protein load as long as 
the protein valency is not too high, as discussed in Appendix~A.
For our system with   lysozyme (surface charge density $\simeq$0.17e/nm$^2$) in 
monovalent salt and the  weakly charged gel (charge fraction 1/10), Debye-H\"uckel level 
assumptions are justified.

\subsection{Osmotic and elastic swelling equilibrium}

In equilibrium the gel size is determined by a mechanical balance between osmotic and elastic 
pressures in the gel via~\cite{flory,dubrovskii, Horkay, Rubinstein:Colby, buschmann, Overbeek} 
\begin{eqnarray}
\pi = \pi_{osm}+\pi_{el} = 0. 
\end{eqnarray}
The osmotic term $\pi_{osm}$ has two contributions one from the solvent, the other from the ions.
The first is often expressed by a de Gennes-like scaling type of relation $\propto V^{-n}$, 
where typically $n=9/4$ for neutral polymers under good solvent 
conditions.~\cite{degennes, Horkay} For charged networks corrections may arise,~\cite{Rubinstein:Colby} 
while for our case of weakly charged and weakly stretched gels the 9/4 exponent 
likely to be valid.~\cite{Horkay, hu:jcp, dubrovskii} 
(According to definitions by Dobrynin and Rubinstein,~\cite{dobrynin} we estimate an elecrostatic blob size of about 1.3 nm using a monomer length 0.35 nm and our charge fraction 1/10. This is indeed smaller than the correlation length of 
3.2 nm given by  our mean mesh  size, estimated from the linker density.)

The ionic osmotic pressure is dominated by the ideal gas pressure of ions in the
gel as given by eq.~(\ref{ionpressure}). This is counterbalanced by the elastic pressure $\pi_{el}$ 
as represented essentially by the shear 
modulus $G$. Weakly charged networks obey Gaussian chain 
statistics~\cite{Rubinstein:Colby, dubrovskii}, and $G$ scales via the power law $\pi_{el}\propto V^{-m}$ with 
$m=1/3$.  Various experiments corroborate that scaling.~\cite{Horkay, hu:jcp, skouri:macromolecules, nisato, dubrovskii, dubrovskii:rakova}
The expression for the total pressure is
\begin{eqnarray}
\label{pressure_balance}
\pi &=& AV^{-n}+BV^{-m}+\pi_{ion}(y)\\ \nonumber  
&=& AV^{-n}\left[ 1-\left(\frac{V}{V_0}\right)^{n-m}\right]+ \pi_{ion}(y),   
\end{eqnarray}
where $A$ and $B$ (or $V_0$) are volume-independent constants.  For high salt concentration the 
ionic contribution vanishes, $\pi_{ion}=0$,  and the equilibrium gel is in the 'neutral' reference
state with volume $V_0$. The bulk modulus of the gel in this state is defined via $\beta K_m(V_0) = -V\partial \beta P/\partial V|_{T,V_0} = (23/12)A/V_0^n$.
By fitting eq.~(\ref{pressure_balance}) to a variety of salt concentrations (without proteins) 
we will obtain both the unknowns $A$ and $V_0$ and thus $K_m(V_0)$. 

In first order the effect of protein addition to the gel is lowering of the gel net charge 
and the inhomogenization of the charge and electrostatic potential distribution inside the gel. 
As shown in the PB cell model in Appendix A  those effects can be approximately included in (\ref{pressure_balance}) by 
replacing $\pi_{ion}(y)$ with 
\begin{eqnarray}
\pi_{ion}^p(y) \simeq \pi_{ion}(y)  +2k_BT c_s[\tilde\phi(R_c)-\Delta\tilde\phi]\sinh(\Delta\tilde\phi), 
\label{press_electro}
\end{eqnarray}
with $y = (z_g c_g + z_pN_{bp}/V)/(2c_s)$. The potential 
$\tilde\phi(R_c)$ is the electrostatic potential at the cell boundary $R_c$ in the PB cell model (Appendix A). 

Estimating other protein-induced contributions to the pressure from more local effects on elasticity, such 
as cross-linking by local binding,~\cite{borrega}  conformational restraints of the polymer network, 
or possible contributions from the protein osmotic pressure is challenging due to the lack of 
precise knowledge of the nature of the bound state and is out of scope of this paper. 

\subsection{Binding isotherms}

The transfer of one protein into the gel will be accompanied by a release of the binding  
enthalpy $\Delta H$. If $N_{bp}$ proteins bind and we assume that the heat per protein 
does not change with load, the total  heat released is 
\begin{eqnarray}
Q(N_{bp}) = \Delta H c_mV_{\rm tot} N_{bp},  
\end{eqnarray}
where $V_{tot}$ is the total titration volume. In the ITC experiments $Q$ is measured vs. the total protein concentration $c_p^{tot}$. Introducing the molar ratio $x = c_p^{tot}/c_m$ we can write $Q(x) = \Delta H c_m V_{\rm tot} N_{bp}(x)$.  
The incremental heat $Q'(x)=\partial Q/\partial x$ per molar concentration of protein is more sensitive to 
fitting and reads
\begin{eqnarray}
Q'(x)/(V_{tot}c_p^{tot}) = \Delta H N_{bp}'(x)/x.
\label{Qprime}
\end{eqnarray}
For large binding constants and small $x$, almost all of the proteins immediately get sorbed, and 
$N_{bp}(x)\simeq x$, and a plateau of heigh $Q'(x)/(V_{tot}c_p^{tot})\simeq \Delta H$ is expected, 
independent of the binding constant. For large $x$, typically $N_{bp}(x)$ saturates and $Q'(x)\propto N_{bp}'(x)=0$. 
Thus we recognize that fitting to binding models is most sensitive to intermediate values of the molar ratio $x$, 
in the pre-saturation regime, near the inflection point of $Q'(x)$.

\subsubsection{Standard Langmuir Binding Model}

The standard Langmuir model is based on identifying the association reaction A+B$\rightarrow$AB, with 
a binding constant K = [AB]/[A][B], where the square brackets denote concentrations. 
The basic assumptions in the Langmuir model are that ideal particles are binding to a fixed number $N$ of identical and independent bindings sites.  
The equilibrium constant in the Langmuir  framework  is usually written as~\cite{langmuir,langmuir2}
\begin{eqnarray}
K = \frac{\Theta}{(1-\Theta)c_p}, 
\label{langmuir}
\end{eqnarray}
where $\Theta=N_{bp}/N$ denotes the fraction of bound protein $N_{bp}$ and total sites $N$ (see also 
derivation in Appendix B for the canonical ensemble).
The protein density outside the gel can be expressed
by the total protein density in the sample minus bound protein density by  $c_p = c_p^{\rm tot} - N\Theta c_m$. 
If it is assumed that $N$, $K$,  and $\Delta H$ are protein concentration independent, solving (\ref{langmuir}) for $\Theta$ 
gives the total heat
\begin{eqnarray}
Q(x)  = \frac{1}{2}{N\Delta H c_mV_{\rm tot}}\left [\xi - \sqrt{\xi^2-{4x}/{N}} \right], 
\label{langmuir2}
\end{eqnarray} 
with $\xi = 1+ x/N + 1/(NKc_m)$.
The fitting of $Q'(x)/(V_{tot}c_p^{tot})$ to the experimental data then yields the unknown 
constants $K$, $N$, and $\Delta H$. 
Typically a sigmoidal curve is measured for $\tilde Q'(x)$, where $\Delta H$ describes the plateau for the first injections (small $x$),  $N$ the inflection point, and $K$ the sharpness of the transition at $x=N$.  Fitting to Langmuir isotherms 
is thus most sensitive close to $x\simeq N$, when the molar ratio equals the number of available
binding sites. 

The Gibbs binding free energy in the 
Langmuir model is defined by 
\begin{eqnarray}
\beta \Delta G = -\ln (K/v_0), 
\end{eqnarray}
where $v_0$ is  the 'standard volume' which describes the 'effective' configurational volume in one of
the binding boxes (see also Appendix B).
Thus the absolute value of $\Delta G$ depends on the magnitude of the standard volume $v_0$, an often
overlooked fact in literature.~\cite{zhou} Typically $v_0 = {\rm l/mol}\simeq 1.6$~nm$^3$ is chosen which is a 
reasonable choice for molecular binding where spatial fluctuations are on a nanometer length scale.
While for quantitative estimates thus precise knowledge of the nature of the bound state is necessary, 
we will satisfy ourselves in this work with the standard choice $v_0 = {\rm l/mol}$.

\subsubsection{Extended Langmuir Binding: Separating Nonelectrostatic and Electrostatic Contributions}

When many charged entities bind to charged regions a cooperativity effect comes
into play due to the change of the global electrostatic properties during loading.  This has been appreciated 
in the Guoy-Chapman-Stern theory for the binding of charged ligands to charged surfaces.~\cite{haydon,seelig,mclaughlin}  Consequently, the total binding constant $K=K(x)$ has to be defined more generally and split up into 
an 'intrinsic' part $K_0$ and an electrostatic  part via 
\begin{eqnarray}
K(x) = K_0\exp[-\beta \Delta G_{\rm el}(x)] = \frac{\Theta(x)}{[1-\Theta(x)]c_p},  
\label{extendedL}
\end{eqnarray}
i.e., $\Delta G(x) = \Delta G_{el}(x)+\Delta G_0$.
In our work we assume $\Delta G_{\rm el}(x)$ to be given by eq.~(\ref{transfer}). 
The intrinsic, $x$-independent binding constant $K_0$ defines the intrinsic 
adsorption free energy 
\begin{eqnarray} 
\beta \Delta G_0 = -\ln (K_0/v_0),
\end{eqnarray}
which only contains contributions from specific interaction between the protein and the gel environment, and
the nonspecific, leading order electrostatic effect has been separated out.  
Specific interaction may include local solvation effects -- where the hydrophobic effect is naturally one  of 
the biggest contributors -- and possibly specific local interactions, such as salt bridges.  

\subsubsection{'Excluded Volume' (EV) Binding Isotherms}

An alternative binding isotherm that may be used for evaluating  the ITC data is based on the equivalence of
chemical potentials in bulk and inside the gel, where in addition to electrostatic and intrinsic 
binding effects the EV interaction between the (hard spherical) proteins 
inside the gel are taken into account.~\cite{biesheuvel:pre,vos:langmuir,cellmodel:wittemann,hansson} 
One assumption is here that in the very contrast to Langmuir models, 
the proteins are not 'condensed' to fixed sites in the gel but are able to freely move around, 
under the restraint only that their translational freedom is confined by packing 
(excluded volume). This ansatz yields the Boltzmann-like equation for the density 
of hard spheres in the gel  
\begin{eqnarray}
N_{bp}/V = \zeta' c_p \exp(-\beta \Delta G_{el}-\beta \Delta G_0)\exp(-\beta \mu_{CS}), 
\label{EV}
\end{eqnarray}
where  $\zeta'$ is the partition function of the bound state more precisely defined later, and $\mu_{CS}$ is the Carnahan-Starling (CS) excess chemical potential~\cite{hansen:mcdonald}  
\begin{eqnarray}
\beta \mu_{CS}=\frac{8\eta-9\eta^2+3\eta^3}{(1-\eta)^3} 
\label{mucs}
\end{eqnarray}
representing the free energy of transferring one hard sphere to a
solution with the packing fraction $\eta = (N_{bp}/V) \pi\sigma_p^3/6$. 
The effective diameter of the sphere 
$\sigma_p=2R_p$ is expected to be close to the diameter of gyration or  hydrodynamic diameter
of the protein (i.e., $\sigma_p \simeq 3-4$~nm for lysozyme)~\cite{hamill} and 
will serve  as a  fitting parameter in the following.   Since packing effects play a role
in the EV approach we correct the gel volume by the polymer volume 
fraction.  A PNIPAM monomer has an excluded volume
roughly of $0.3$~nm$^3$. We estimate $3.7\cdot$10$^6$ monomers in
one microgel what makes $V_{\rm polymer} \simeq 1.1\cdot$10$^6$~nm$^3$. 
This  corresponds to a polymer volume fraction in the range of 5\% (pure, unloaded gel) to 12\% (fully loaded 
gel in 7mM salt).

In (\ref{EV}) we have also separately 
described the nonspecific electrostatic part ($\Delta G_{el}$) and the intrinsic 
part ($\Delta G_0$) to binding as  in the Langmuir model above.
In the EV model, the contribution to $\Delta G_0$ can be still 
thought of being induced by hydrophobic interactions or salt bridges but only in a weak-interaction sense, 
such that the particles in the gel are still translationally free. However, we additionally consider $\zeta'$ in 
eq.~(\ref{EV}), 
the partition function of the protein in the bound state, which can include contributions 
from vibrational and orientational restrictions of the protein's degrees of freedom 
within the gel, e.g., by partial sliding on polyelectrolyte chains.~\cite{henzler:prl}

\subsubsection{Equivalence Between the Langmuir and the EV Approach} 
\label{equiv}

For small protein packing fractions we may linearize the CS chemical potential in eq.~(\ref{EV}).  
We obtain 
\begin{eqnarray}
\frac{N_{bp}}{V} \simeq \zeta' c_p \exp(-\beta \Delta G_{el}-\beta \Delta G_0)\left(1-{2B_2N_{bp}}/{V}\right),
\end{eqnarray}
where we identified the second virial coefficient of hard spheres $B_2=2\pi\sigma_p^3/3$. Physically $2B_2$ 
describes the volume excluded to the centers of the other spheres taken 
by one sphere.   We rearrange to obtain
\begin{eqnarray}
\exp(-\beta \Delta G_{el}-\beta \Delta G_0) = \frac{N_{bp}/V}{(1-{2B_2N_{bp}}/{V})c_p}. 
\end{eqnarray}
If we now make the substitution $N = V/(2B_2)$ and $\Theta = N_{bp}/N$ as
in the Langmuir model above, we end up with the standard Langmuir form (\ref{langmuir})
\begin{eqnarray}
\zeta' 2B_2 \exp(-\beta \Delta G_{el}-\beta \Delta G_0) = \frac{\Theta}{(1-\Theta)c_p}. 
\end{eqnarray}
If $\zeta'=1$, thus the EV treatment is equivalent to  a 
Langmuir picture where a bound ideal gas 
particle has a configurational freedom (volume) of $2B_2$, i.e., it can freely move
around in one of the $N$ binding boxes.  However, as discussed above, in the Langmuir-type 
bound state the configurations are restricted to an effective configurational volume $v_0$ with 
respect to $2B_2$, such that the partition function $\zeta' = v_0/(2B_2)$. With that definition 
we exactly end up with the standard Langmuir model eq.~(\ref{langmuir}).

Thus, in the approximation of small protein packing $\eta\ll1$, the EV 
approach and the standard Langmuir model, eq.~(\ref{langmuir}),  
are formally equivalent and are allowed to be compared, if $N = V/(2B_2)$ in the 
Langmuir picture and  $\zeta' = v_0/(2B_2)$. The parameter $N$, the number of fixed binding sites,
can then be interpreted as the maximum number of binding spots available for hard spheres 
simply due to packing in the available volume $V$.  In both models, Langmuir and EV, 
the binding constant is referenced with respect to a  standard 
volume $v_0 = {\rm l/mol}$.

\subsection{Numerical evaluation including volume change}

In the Langmuir approach the derivative $Q'(x) = \partial Q(x)/\partial x$ of 
eq.~(\ref{langmuir2}) is fitted to the experimental data by scanning through
$\Delta H$, $K_0$, and $N$ values until the least square deviation (LSD) to the 
experimental  data is minimized. The radius $R_p$ is fixed in the Langmuir fittings 
to $1.8$~nm. In the EV approach, eq.~(\ref{EV}) is solved numerically 
and $Q'(x)$, as more generally defined in eq.~(\ref{Qprime}) is fitted to the 
experimental data by minimizing the LSD. Here the fitting parameters are  
$\Delta H$, $K_0$, and $R_p$. 

One challenge arises because also the gel volume $V=V(x)$ is in general 
a function of $x$, the protein concentration. The electrostatic transfer energy 
defined by  eq.~(\ref{transfer}), in turn, depends on the gel volume.  
Thus, for separating out the electrostatic contributions to binding, the fitting of
binding isotherms needs take into account the volume change. 
Since a predictive theory for $V(x)$ is 
out of scope of this paper, we satisfy ourselves in the following by employing 
experimental DLS data for $R(c_s)$ and $R(x)$ at 7~mM salt concentration 
as shown in Fig.~2.  We fit 
$V(x) = 4\pi[R(x)^3-R_{\rm core}^3]/3$ by the empirical function
\begin{eqnarray}
R(x) = \frac{1}{2}(R_{\rm max}-R_{\rm min})\left[1- {\rm tanh}\frac{x-x_0}{\Delta} \right]+R_{\rm min}, 
\label{tanh}
\end{eqnarray}
with $R_{\rm max}$ and $R_{\rm min}$ being the maximum and minimum gel 
radius at $x=0$ and $x\rightarrow\infty$, respectively, $x_0$ is the location
of the inflection point, and $\Delta$ the distribution width. For $T=298$~K and 7~mM salt  
we find $R_{\rm max}=172$~nm, $R_{\rm min}= 129$~nm,  $\Delta = 24000$, 
and $x_0$ can be identified with $N({\rm 7 mM})$. For the other salt concentrations, where no DLS data are
available, $V(x)$ is obtained by using the known $R_{max} = 165.7$~nm (17 mM) and 157.1~nm (32 mM) 
(cf. Fig. 2a) and $x_0=N$ (cf. Tables II-IV), while  $R_{\rm min}=129$~nm is used from the 7~mM fit. 
The now only free variable $\Delta$ is employed as an additional parameter obtained by least 
square fitting the ITC data.  We find $\Delta$ = 31000 ($c_s=$~17~mM) and 42000 ($cs=$~32~mM) 
which appears  correlated with the change of the sharpness of the binding isotherms $N_{bp}(x)$ with $c_s$.

\section{Results and Discussion}

\subsection{Gel shrinking by salt and proteins} 

In agreement with previous observations,~\cite{eichenbaum,kabanov, hansson} DLS shows that the charged 
PNIPAM microgel shrinks upon the systematic addition 
of salt  as summarized in Fig.~2 (a)  up to a concentration of $c_s= 1$~M. The gel
radius decreases monotonically and saturates at molar concentrations at a radius close to $R \simeq 139$~nm.  The best fit of 
our mechanical balance approach eq.~(\ref{pressure_balance}), also shown in Fig.~2 (a), yields very good agreement 
in almost the whole range of $c_s$.  From the fit 
we obtain $R(c_s\rightarrow\infty)=R_0 = 138.5$~nm and a bulk modulus $K_m(V_0) =  204$~kPa. 
Assuming Poisson's ratio to be 1/3,  
valid for neutral polyacrylamide or poly-NiPAm gels,~\cite{poisson1,poisson2} then Young's modulus $E\simeq K$. 
Our determined value for $K$ is thus fully consistent with recently measured Young moduli 
investigated for varying BIS cross linker density at similar temperatures.~\cite{klitzing} In the latter $E$ was 
measured between $\simeq80$~kPa and  $\simeq 500$~kPa for BIS contents of 2\% and 10\%, respectively, 
compared to  5\% in this work. The good agreement supports our considerations leading
to eq.~(\ref{pressure_balance}).

\vspace{1.0cm}
\begin{figure}[h]
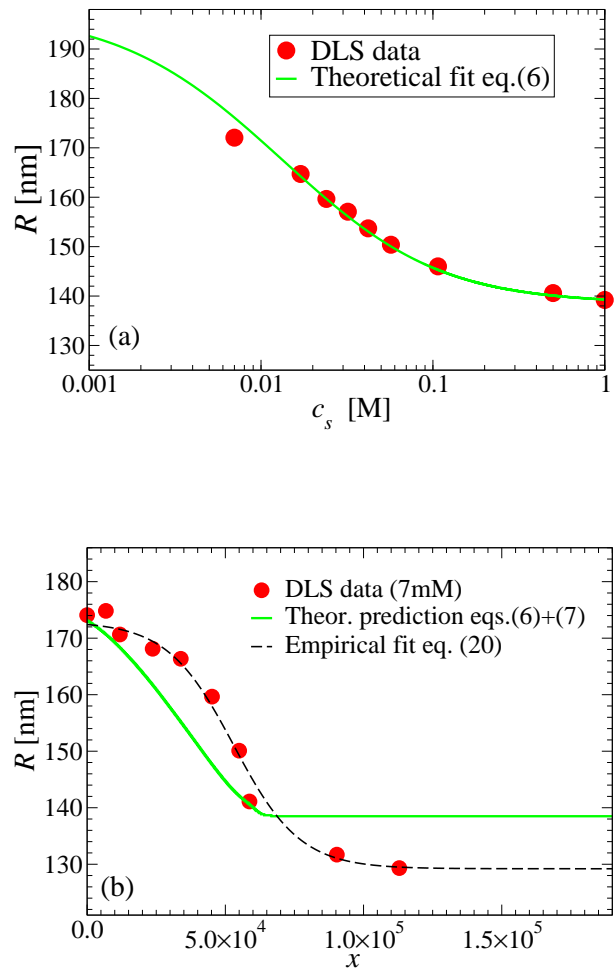

\centerline{\includegraphics[width=8cm,angle=0]{fig2a.eps}}
\vspace{1.5cm}
\centerline{\includegraphics[width=8cm,angle=0]{fig2b.eps}}
\caption{(a) Microgel radius $R(c_s)$ vs. bulk salt concentration $c_s$ (no proteins). Experimental DLS data (filled circles) 
are fitted by eq.~(6) finding $R_0=138.5$~nm and a bulk modulus $K\simeq 204$~kPa.  (b) Microgel radius $R(x)$ 
vs. lysozyme molar ratio $x=c_p^{\rm tot}/c_m$. The green solid line is the theoretical prediction from eqs.~(6) and (7) 
based on electrostatic considerations only. The dashed line depicts a simple empirical fit to $R(x) = \frac{1}{2}(R_{\rm max}-R_{\rm min})\left[1- {\rm tanh}\frac{x-x_0}{\Delta} \right]+R_{\rm min}, 
$ [eq.~(20)] used as input
for the fitting of ITC binding isotherms. }
\label{gelradius}
\end{figure}

In Fig. 2(b) we show the measurement of the hydrogel radius $R(x)$ vs. the molar ratio $x$ 
for the system at 7~mM salt concentration. Analogously to the salt-only case, shrinking of the gel is 
now observed with increasing protein concentration, i.e., with increasing protein load. 
At the highest investigated ratio $x\simeq 1.1\cdot 10^5$ the radius is notably smaller than $R_0$ 
pointing to binding-related network tightening.  By fitting to the  the empirical 
function~(\ref{tanh}), we find $R_{\rm min}= 129$~nm. Thus the gel volume under full protein 
load is $V_0/V_{min}\simeq 1.3$ times smaller than in the neutral reference state. If we assume that the 
charged proteins under full load lead to complete charge neutralization and thus $\pi_{ion}\simeq 0$, then
the bulk modulus induced by protein binding scales  $\propto (V_0/V_{\rm min})^{9/4}$ and is $\simeq 1.7$ 
times larger than without proteins, i.e., the gel is roughly 2 times stiffer due to protein sorption. This has to be considered a 
lower bound as it is  likely that ionic contributions to the  osmotic  pressure still play a role, and
maybe protein osmotic effects due to EV interactions need to be considered. 

However, in order to check to what amount purely electrostatic effects by proteins 
induce gel shrinking, we plot the description by eqs. (\ref{pressure_balance}) and (\ref{press_electro}) 
also in Fig. 2 (b) using the experimental ITC binding isotherm $N_{bp}(x)$ as input.  
This description is now a prediction and no fitting is involved. While the overall shrinking of the gel is 
reasonably captured, the model yields a too fast decrease for small $x$. This may point
to shortcomings of the mean-field cell model (Appendix A), where no ion and protein fluctuations are included. 
For large $x$, the experimental saturation of $R_{min}\simeq 129$~nm is not reached indicating that 
nonelectrostatic effects to gel elasticity play a role. Similar unsatisfying performances of
simple Donnan models have been observed also in a recent work.~\cite{hansson} 
However, from our comparison it is quite reasonable to judge that the dominant effect to 
gel shrinking by protein uptake is of ion osmotic origin. 

\subsection{Characterizing experimental binding isotherms}

\subsubsection{Langmuir models}

The evaluation of the ITC data by the Langmuir-type binding models is presented in Fig.~3~(a). 
First we compare the results of different model assumptions for 7~mM 
ionic strength: i) standard Langmuir model vs. ii) extended Langmuir (including electrostatic 
cooperativity) with constant volume vs. iii) extended Langmuir including the DLS-measured 
volume change. From looking at the fits by eye and judging
from the overall least square deviation (LSD) to the ITC data, cf. Tab.~II, all fits look comparably well 
and can only be distinguished in the presaturation region around $x\simeq 5\cdot10^4$. 
As discussed above fitting is most sensitive to this transition region. The fitting 
parameters are summarized in Tab.~II: while the heat of binding $\Delta H\simeq 60$~kJ/mol 
and total number of binding sites $N\simeq 63000$ are relatively insensitive to the model assumptions, 
the changes in the binding constant are big.  From the standard model 
$\Delta G \equiv \Delta G_0 = -\beta\ln(K/v_0) = -36.6$~kJ/mol.
In  the extended model without volume change ii),  $\Delta G_0 = -24.5$~kJ/mol., i.e., 
more than 12  kJ/mol correspond to $x$-dependent electrostatic contributions. 
Including the volume change, however,  has a considerable effect on the electrostatic contribution 
which grows by 18 kJ/mol.  The latter trend is understood by the fact
that the monomer charge density $c_g=N_g/V$ increases with shrinking and the 
contributions in  $\Delta G_{el}$ as given by eq.~(\ref{transfer}) rise. 
Thus considering volume changes in charged systems is important for 
quantitative fitting, especially in those systems where deswelling is significant.  
The value of the intrinsic binding free energy is $\Delta G_0 = -18.3$~kJ/mol 
for 7~mM ionic strength.

In the next step the extended Langmuir model including volume change  has been applied
to the other ionic strengths 17 and 32~mM as also shown in Fig. 3(a) and summarized in Tab.~II.  
Note again that the starting gel volumes at $x=0$ decrease for increasing ionic strengths, see the 
values for $R_{max}$ in Tab.~II. Here we first observe that the number 
of Langmuir binding sites $N$ decreases with higher ionic strength. The reason is {\it a priori} unclear 
as the Langmuir model assumes a fixed number of binding sites independent of ionic strength.  
We further notice that the heat of adsorption $\Delta H$ slightly increases with ionic strength. 
More importantly, however, $\Delta G_0$ stays fairly independent of $c_s$.  Thus,  in contrast to the 
standard Langmuir model, the nonspecific  electrostatic  contributions have been consistently 
separated out and the remaining $\Delta G_0$ becomes salt concentration independent. 
On average we find $\Delta G_0 \simeq -18$~kJ/mol  that might be attributable to hydrophobic 
interactions or possibly other local binding effects.

\vspace{1.0cm}
\begin{figure}[h]
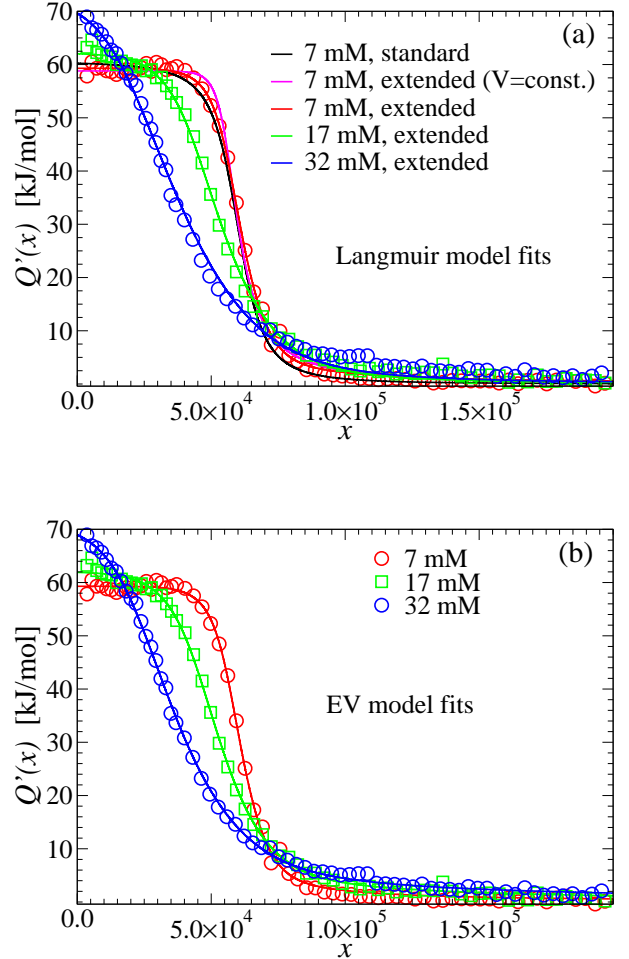

\centerline{\includegraphics[width=8cm,angle=0]{fig3a.eps}}
\vspace{1.0cm}
\centerline{\includegraphics[width=8cm,angle=0]{fig3b.eps}}

\caption{Fitting of the experimental ITC data (symbols) at 7, 17, 32 mM salt concentrations by different binding models. 
(a) Langmuir fitting. For 7~mM ionic strength we compare the standard Langmuir description (black line), to the Langmuir model including electrostatic cooperativity with constant gel volume (magenta) and changing gel volume according to 
eq.~(\ref{tanh}) (red). The data for 17 (green) and 32 mM (blue) are fitted using the extended model including the volume change. See Table~II for obtained fitting parameters. (b) Application of the Excluded Volume (EV) binding model to 
the same ITC data as in (a). See Table~III for obtained fitting parameters. }
\label{bla}
\end{figure}

\begin{table}[h]
\begin{center}
\begin{tabular}{|c||c|c|c|c|c|c|}
\hline
$c_s$  &  $ R_{\rm max}$ & $N$  & $\Delta H$  & $K_0$ & $\Delta G_0$ & LSD \\
  $\rm [mM]$ &   [nm]                   &         & [kJ/mol]      & [l/mol]  & [kJ/mol]          &    \\
\hline
7  (i)                         & --        & 60100  & 61  & 2.6$\cdot10^{6}$  & -36.6  & 132 \\
\hline
7  (ii)   & 172.1 & 65500  & 59  & 1.9$\cdot10^{4}$ & -24.5  & 89  \\
\hline
7     			  & 172.1 & 63000  &  59 & 1.6$\cdot10^{3}$  & -18.3  &44  \\
17                        & 165.7 & 57500  & 62  & 2.0$\cdot10^{3}$ & -18.9  &57  \\
32                       & 157.1 & 42500  & 72  & 1.3$\cdot10^{3}$ & -17.8  &130  \\
\hline
\end{tabular}
\end{center}
\caption{Results of fitting to Langmuir models at $T=298$~K. In the 1st row (i) the results of the 
standard Langmuir fit without any correction is shown. In the 2nd row (ii) the standard model 
was corrected for electrostatic cooperativity with a constant gel volume. In rows 3 to 5 the change of gel volume was
considered in the fit according to eq.~(\ref{tanh}). LSD denotes the least square deviation to the experimental data.
Lower values correspond to better quality of the fits.}
\label{table2}
\end{table}

\subsubsection{Excluded Volume model}

The fitting of the ITC data by the EV model is presented in Fig.~3~(b).
Corresponding fitting parameters $\Delta H$, $R_p$, and $\Delta G_0$ 
are summarized in Tab.~III. As in the Langmuir models the binding isotherms 
are described very well by the fits judging from the overall 
LSD to the experimental data. The results for the heat $\Delta H$ are also consistent
with the Langmuir fits. This is not surprising as this value is determined
by the plateau in $Q'(x)$ for small $x$ far away from the saturation 
regime. The fitting parameter $N$ is now replaced in the EV model by the
parameter $R_p$, the effective hard-core radius of the protein. The analysis 
yields values of $R_p$ between $1.8$ to $2.0$~nm, see Tab. III, 
weakly  depending on ionic strength. Those numbers are indeed very 
close to the hydrodynamic radius of lysozyme of about $1.7$~nm.~\cite{hamill}  
This good agreement is actually remarkable and justifies the assumptions
leading to the EV model, i.e. corroborates well with a packing picture of
globular proteins.

The magnitude of the intrinsic adsorption energy $|\Delta G_0|$ in Tab.~III 
are between 15 and 18 kJ/mol matching closely the ones from the Langmuir model 
in Tab.~II. A very small salt dependence of $\Delta G_0$ 
remains indicating a slightly less accurate subtraction of the nonspecific effects
in this model. However, the salt concentration dependence is
quite small and on average we find $\Delta G_0 \simeq -17$~kJ/mol 
that has to be attributed to specific local binding effects. In Tab.~III we also show the 
results of EV model fitting to experimental  data gathered at a higher temperature, 
$T=303$~K. The intrinsic binding affinity increases slightly to an average 
$\Delta G_0 \simeq -18$~kJ/mol which corroborate with  the typical 
thermodynamic signature of  increasing hydrophobic association 
at enhanced temperatures.~\cite{dill}

\begin{table}[ht]
\begin{center}
\begin{tabular}{|c||c|c|c|c|c||c|}
\hline
$c_s$ & $ R_{\rm max}$ & $R_p$ & $\Delta H$ & $\Delta G_0$ & LSD & $T$\\
$\rm [mM] $&  [nm] & [nm] & [kJ/mol] & [kJ/mol] & & [K] \\
\hline
7     			  & 172.1 & 1.77  &  59  & -14.9 & 30 &  \\
17                        & 165.7 & 1.83  & 62     & -17.9 & 33 & 298 \\
32                       & 157.1  &  2.01  & 71   & -18.4   & 25 & \\
\hline
7     			  & 172.1 & 1.78  &  70    & -15.1 & 111 & \\
17                        & 165.7 & 1.79  &  70    & -18.9 & 43  & 303\\
32                       & 157.1  &  1.85  & 65   & -19.6   & 75  &\\

\hline
\end{tabular}
\end{center}
\caption{Results of fitting to the Excluded Volume (EV) model at $T=298$~K (top) and $T=303$~K (bottom).  The change of gel volume was considered in the fit according to eq.~(\ref{tanh}). LSD denotes the least square deviation to the experimental data.
Lower values correspond to better quality of the fits.}
\label{table2}
\end{table}

\subsubsection{Donnan potentials $\Delta\phi(x)$ and the binding constant $K(x)$}

\begin{figure}[ht]
\vspace{1.0cm}
\centerline{\includegraphics[width=8cm,angle=0]{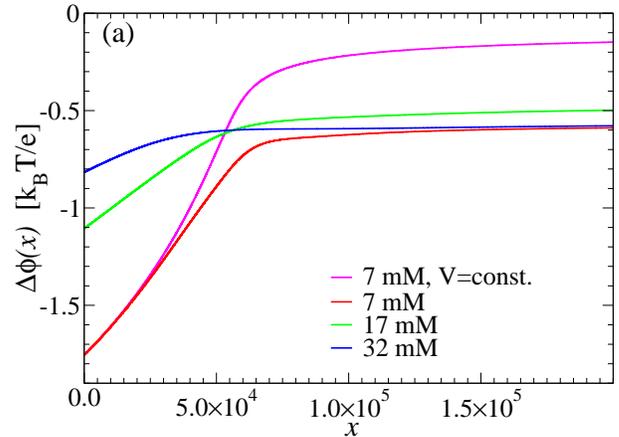}}
\vspace{1cm}
\centerline{\includegraphics[width=8cm,angle=0]{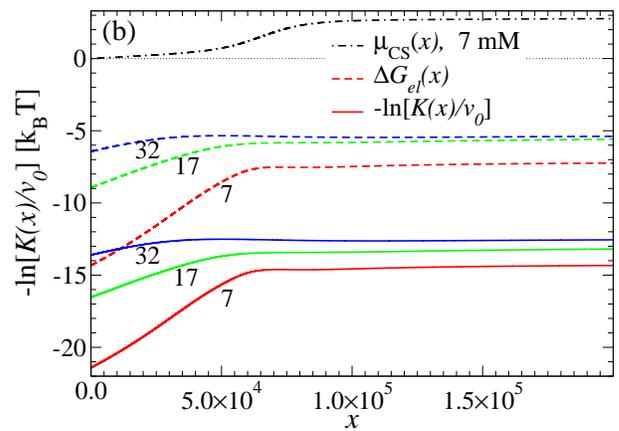}}
\caption{(a) Mean (Donnan) potential $\Delta\phi(x)$ inside the gel vs. molar ratio $x$ 
according to the Donnan form eq.~(1) and (4)
obtained from fitting of the extended Langmuir model to the ITC data. For 7~mM we compare the results from fitting with constant volume vs. variable volume. 
(b) Total binding affinities $-\ln K(x)/v_0$ (solid lines) and total electrostatic contribution $\Delta G_{el}(x)$ (dashed lines), 
cf. eq.~(\ref{transfer}), to the protein transfer energy from the extended 
Langmuir model for the different ionic strengths 7, 17, 32 mM, as labeled. Also plotted is the hard-sphere 
Carnahan-Starling contribution $\mu_{CS}$ from the EV model fit at 7~mM (dashed-dotted line).}
\label{bla}
\end{figure}

The Donnan potential $\Delta\phi(x)$, the electrostatic binding energy $\Delta G_{el}(x)$, 
 and so the total binding constant $K(x)$ depend on protein load, and thus, in turn, 
  on the molar ratio $x$. The Donnan potential  $\Delta \phi(x)$ as resulting from
  the fitting to the extended Langmuir model is plotted 
in Fig.~4(a) for all ionic strengths.  In the case of $c_s=$~7~mM, at the beginning of the 
titration, $x\simeq 0$, the potential magnitude is
a considerable 1.77 $k_BT$/e ($\simeq 44$~mV).  
The potential 
quickly decreases with protein load due to charge neutralization, c.f. 
eq.~(\ref{ratio}), and saturates for $x\gtrsim N$ due to saturation of $N_{bp}(x)$.   
The decrease of gel 
volume in the adsorption process has a notable effect on the potential at large $x$ 
due to the accompanying increase in monomer charge density. Increasing the ionic 
strength, on the other hand, considerably lowers the potential at small load, while surprisingly
$\Delta\phi$ remains roughly independent of $c_s$ for high load. The reason for the 
latter is that less proteins bind for higher ionic strengths, i.e., there is
less charge neutralization. The Donnan potential is expected to be 
highly correlated with the surface potentials of the microgels which 
govern colloidal stability in solution.~\cite{hansson}

The total electrostatic contribution $\Delta G_{el}(x)$ to the binding constant 
is plotted in Fig. 4(b) together with the total binding free energy $-k_BT\ln[K(x)/v_0]$. 
The curves $\Delta G_{el}(x)$ and $-k_BT\ln[K(x)/v_0]$ 
are parallel with an offset given by the intrinsic part $\Delta G_0$, 
as defined in eq.~(\ref{extendedL}).  The electrostatic contribution for very small load ($x=0$)
is big and contributing a favorable 15~$k_BT$ ($\simeq 37$~kJ/mol) for 7~mM and 6~$k_BT$ 
($\simeq 15$~kJ/mol) for 32~mM to the total binding affinity of about 22 $k_BT$ and 14 $k_BT$, 
respectively, at the beginning of the titration. We find that  the Born contribution (second term in 
(\ref{transfer})) constitutes a significant (favorable) 2-3 $k_BT$ to the total electrostatic energy and is 
thus not negligible. All curves decrease in their absolute values 
and saturate for values $x\gtrsim N$, when the molar ratio exceeds the number
of binding sites in the Langmuir model. Note that for 7~mM the saturation 
value of $K(x)$ is close to the value of $K$ obtained from standard Langmuir fitting, c.f., 
Tab.~II, because fitting is most sensitive to the presaturation region $x \simeq N$ 
above which $K(x)$ remains constant. This statement holds also for the other
salt concentrations when $K(x)$ is compared to previous standard Langmuir 
fitting.~\cite{nicole}

Also shown in Fig.~4~(b) is the Carnahan-Starling contribution $\mu_{CS}$ in 
the EV model. This entropic penalty due to hard sphere packing slowly rises
with $x$ in the regime $x\lesssim N$ until quickly increasing to unfavorable 3 $k_BT$ in 
the saturation regime  $x\gtrsim N$. 

\subsection{Interpretation of Langmuir vs. EV model results}

In section \ref{equiv} we argued that the EV model is equivalent to the Langmuir approach 
in the low packing regime ($\eta\ll1$), if $N$, the number of binding sites, is  equated
to $V/(2 B_2)$ the number of free spots just by packing.  Indeed
if one now considers a mean gel radius of $R = 150$~nm and protein diameter 
$\sigma_p=3.8$~nm obtained from fitting above, we end up with 
$N =  V/(2B_2) \simeq 59000$,  totally consistent with numbers obtained from 
fitting to the Langmuir models above, cf. Table II.  This agreement implies
that the strict Langmuir assumption of a fixed set of binding sites can still
be considered an interpretable quantity, even if the nature of the bound
state is ill-defined. From that point of view, the decrease in $N$ in the Langmuir models
with increasing ionic strengths can be understood: for increasing $c_s$,  the gel volume 
$V$ decreases and packing penalties are becoming more important for a  smaller 
number $N_{bp}$ of bound protein. Hence, in standard Langmuir fitting of ITC 
data for sorption to soft materials,~\cite{itc,linse:pnas,linse:nanoletters, rotello,kissel2, nicole} 
where it is likely that not a condensation-like binding of proteins takes place, 
the stochiometry of binding may be interpretable by packing effects. 

As we demonstrated in Fig.~4 the electrostatic energies and therefore the binding affinity $K(x)$
are functions of the molar ratio $x$.  How do we interpret a constant $K$ as deduced
from standard Langmuir fitting?~\cite{itc,linse:pnas,linse:nanoletters, rotello,kissel2, nicole}
As argued in the Methods section fitting is most sensitive to the presaturation 
region $x\simeq N$.  Indeed the data in Fig.~4~(b) shows that $K(x\simeq N)$ equals
the values of standard Langmuir fitting, see Tab.~II. Thus, we can conclude that 
a binding constant obtained from a standard Langmuir fit is a reasonable
number which can be interpreted as  binding affinities in the presaturation regime, 
i.e., in the intermediate to high protein load regime, where also volume changes
are not so large anymore.  However, for small $x\ll N$ (small protein load) our separation into electrostatic 
and hydrophobic contributions shows that binding affinities can be about 1000 
times larger than in the presaturation regime. This may have implications 
for the modeling and interpretation of protein binding kinetics.~\cite{kabanov, linse:kinetics,casals, nicole2}

It may be on first glance remarkable that in both models, extended Langmuir and 
EV, the consistent separation of electrostatic and hydrophobic effects yields the 
same number for the hydrophobic (intrinsic) binding affinity. In our system we find
$\Delta G_0\simeq -7$~$k_BT$ which must be attributed to nonelectrostatic binding
effects and constitutes roughly 1/3 or 1/2 of the total binding affinity in the small
and high load regimes, respectively. The very weak dependence of $\Delta G_0$  
with salt concentration in both models indicates a successful separation of 
nonspecific electrostatic and intrinsic effects in our treatment. However, note
that the agreement can only be established if the nature of the  bound state
is identically defined in both models as discussed in section~\ref{equiv}.
Since at low packing we exemplified total equivalence of both models, 
the agreement for $\Delta G_0$ is thus not unexpected. Probably for 
higher packing densities than observed in this study (where $\eta \lesssim 0.18$), 
the results from both models may disagree stronger. Thus for not too
high packing, one model does not seem superior over the other in describing
experimentally measured binding isotherms, if correctly interpreted.

Furthermore we would like to comment on the magnitude of 
$\Delta G_0\simeq -7$~$k_BT$ for the instrinsic interaction 
of lysozyme with the PNIPAM network. If methane-methane interactions are taken as reference 
with attractions on the order of 2-3~$k_BT$, then 7~$k_BT$
correspond to 2-3 hydrophobic protein-PNIPAM contacts on average which seems reasonable. 
Increasing temperature led to an increase in $\Delta G_0$ conform with the
signature of hydrophobic interactions.~\cite{dill} A recent
study on a similar system showed hardly uptake of 
lysozyme by an uncharged PNIPAM microgel.~\cite{hansson} Reasons for this discrepancy may 
be the different gels preparation state, e.g., larger pore sizes within the gel. 
Alternatively, maybe we overestimate the effects of hydrophobicity and other 
local effects, such as salt bridges, play a more important role as expected.

Finally, we note that effective net charge of chicken egg white lysozyme as used in this work may be 
slightly larger on average due to the protonation effects ($z_p\simeq 7$ to 8) 
within the gel.~\cite{nicole} However, using $z_p$ = +7.5 or +8 in our analysis we end up 
with a similar $\Delta G_0\simeq -7$~$k_BT$; the reason is that while the prefactor in the 
electrostatic contribution (3) rises, the Donnan potential (2) decreases quicker 
with load. These effects roughly cancel each other for our particular system.
\\

\section{Concluding Remarks}

In summary we have systematically introduced binding models to 
characterize the physical interactions in the process of equilibrium protein sorption to microgels. 
The binding models separate out electrostatic cooperativity and
include deswelling effects. The analysis yields a hydrophobic, salt-independent 
interaction of around $\Delta G_0 \simeq -7$~$k_BT$ for our core-shell system
and hen-egg white lysozyme.

The found $\Delta G_0$ constitutes roughly 1/3 or 1/2 of the total binding 
affinity in the small and high load regimes, respectively. For small protein load, binding affinities 
can be about 1000 times larger than in the presaturation regime due to high net charge
of the protein which is gradually neutralized upon binding. This may have implications 
for the modeling of protein binding at early stages of sorption, especially
to understand kinetic rates.~\cite{kabanov, linse:kinetics,casals, nicole2}  Moreover, 
the Donnan potential is expected to be highly correlated with the surface potentials 
of the microgels which govern colloidal microgel stability in solution.~\cite{hansson}

We find gel deswelling to be mostly of electrostatic origin. However, the gel 
becomes at least 2 times stiffer at high load pointing either to more specific 
effects at high load (e.g., cross links) or electrostatic correlations not
accounted for in mean-field PB cell 
model approaches. The change and control of material properties
upon protein load is essential for functionality, ~\cite{levental} and suggest
challenging investigations in future. 

In two complementary models, extended Langmuir and 
EV, the consistent separation of electrostatic and hydrophobic effects yields the 
same number for the hydrophobic (intrinsic) binding affinity.  The agreement can 
only be established if the nature of the  bound state
is identically defined in both models. 
Details of the bound state can be inferred, for instance, from small angle 
scattering.~\cite{henzler:prl} We work at low protein packing; hence, one model does not seem superior over the other in describing experimentally measured binding isotherms, at least for not too high loads (packing). 
With that a more quantitative  interpretation of binding data ~\cite{itc,linse:pnas,linse:nanoletters, rotello,kissel2, nicole, hansson, norde} in terms of separate physical interactions is possible in future. Especially
we have shown that fitting based on standard Langmuir models yields
interpretable binding affinities and stochiometry.~\cite{itc,linse:pnas,linse:nanoletters, rotello,kissel2, nicole}
More challenges arise, however,  if the systems of practical interest exhibit microscopic irreversibiltiy
and the final protein binding cannot be considered in equilibrium terms.~\cite{dawson1,dawson2}

\acknowledgments
Financial support by the Deutsche Forschungsgemeinschaft (DFG), 
Schwerpunkt ÔÔHydrogeleÕÕ, and by the Helmholtz Virtual Institute  is gratefully acknowledged.

\appendix

\section{Linearized Poisson-Boltzmann (LPB) cell model}

To obtain the electrostatic contributions to the transfer of charged hard spheres from 
bulk to gel we make use of the Poisson-Boltzmann cell model. While often numerical solutions have been employed,~\cite{cellmodel:marcus, buschmann,wennerstrom1,wennerstrom2,cellmodel:pincus, barbosa,deserno, allen:warren,hansson:jcis} for weak perturbations the linearized form can be treated analytically.~\cite{cellmodel:denton,cellmodel:wittemann} 

In the cell model each protein is represented by a homogeneously charged sphere with radius $R_p$ 
and valency $z_p$ centered in a spherical cell with radius $R_c$ and volume $V_c$.  
The gel is assumed to be made up of $N_{bp}$ such cells whose dimensions are determined by 
the number of proteins in the volume, i.e., 
\begin{eqnarray}
R_c = \left(\frac{3V_c}{4\pi}\right)^{1/3} = \left(\frac{3V}{4\pi N_{bp}}\right)^{1/3}\propto N_{bp}^{-1/3}, 
\end{eqnarray}
where $N_{bp}$ is the total number of proteins in the volume $V$.  Each cell is in contact with a reservoir 
with monovalent salt concentration $c_s$.  Inside the gel each cell additionally contains a fixed number of 
$N_c = N_{g}/N_{bp}$ charges from the charged network monomers.  The mean monomer 
charge concentration in one cell is  thus $N_c/V_c = N_g/V=c_g$.  Each cell must be electroneutral on average, 
i.e., in the gel it holds
\begin{eqnarray}
c_se^{-\Delta\tilde\phi}-c_s e^{\Delta\tilde\phi}+z_gc_g +z_p N_{bp}/V= 0, 
\label{electroneutral}  
\end{eqnarray}
where we neglected the vanishingly small protein concentration outside of the gel. 
For high protein load, $R_c$ becomes comparable to $R_p$ and the cell volume
needs in principle to be corrected by the protein volume. However, for our systems at highest protein
load $(R_p/R_c)^3 \lesssim 0.1$, and the correction is negligible for small and intermediate loads. 
The solution of (\ref{electroneutral}) is  the modified Donnan potential 
\begin{eqnarray}
\Delta \tilde \phi = \ln[y+\sqrt{y^2+1}]
\label{donnan_mod}
\end{eqnarray}
with $y = (z_g c_g + z_pN_{bp}/V)/(2c_s)$ which describes the difference in the mean electrostatic potential with respect to the bulk reference state where we set $\phi=0$.  In the bulk the protein concentration is 
typically vanishingly small, thus electroneutrality dictates $c_+ = c_- = c_s$ to a very good approximation. 

\begin{figure}[h]
\centerline{\includegraphics[width=9cm,angle=0]{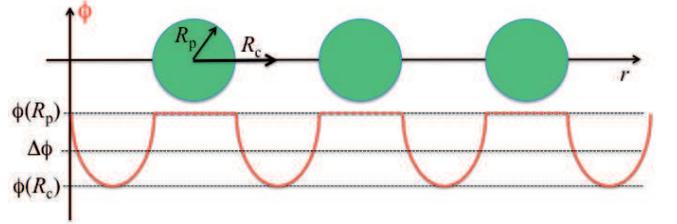}}
\caption{Illustrative 1D sketch of the electrostatic potential distribution in the cell model: the charged proteins (blue spheres) 
with radius $R_p$ sit equidistantly in the gel in their cells with radius $R_c$. $\phi(R_p)$ is the potential
at the protein surface, $\phi(R_c)$ the potential at the cell surface, and $\Delta\phi$ the mean (Donnan) potential. 
In our approach $\phi_1(r)$ is a perturbation of $\phi(r)$ around $\Delta\phi$.}
\label{cellmodel}
\end{figure}

We now focus on the proteins inside the gel. Here we 
assume the PE network in the gel to be flexible and fluid-like and thus the $N_c$ charged monomers to 
behave like mobile counterions to the protein. The PB equation in spherical coordinates is then
\begin{eqnarray}
\label{pb}
\frac{1}{r}(r\tilde\phi)'' &=& -4\pi\l_B [(z_g c\exp(-z_g\tilde\phi_1) \\ \nonumber &-&c_s\exp(\Delta\tilde\phi+\tilde\phi_1)+c_s\exp(-\Delta\tilde\phi-\tilde\phi_1)],   
\end{eqnarray}
where we made the ansatz $\tilde\phi(r) = \Delta\tilde\phi +\tilde\phi_1(r)$, with $\Delta\tilde\phi$ being the 
constant mean (modified 
Donnan)  potential in eq.~(\ref{donnan_mod}) and $\phi_1$ being a perturbation induced by the protein. 
Note that if the monomers were assumed to be just 
a fixed, homogeneous background, they would not couple to the field and $c\exp(-z_g\tilde\phi_1)$ 
needed to be replaced by the fixed concentration $c_g$. The constant $c$ is defined by 
conservation of the number of monomer charges in the cell via
\begin{eqnarray}
N_c = c\int_{V_c} {\rm d}^3r \exp(-z_g\tilde\phi_1).
\label{norm1}
\end{eqnarray}
It also holds that the average potential equals the Donnan potential, i.e,
\begin{eqnarray}
\Delta\tilde\phi = \frac{1}{V_c}\int_{V_c}{\rm d}^3r\, \tilde\phi(r)
\label{norm2}
\end{eqnarray}
For not too large protein charges and thus $\tilde\phi_1 \ll 1$,  we can linearize the exponentials in the 
PB eq. (\ref{pb}) with respect to $\tilde\phi_1$ which yields
\begin{eqnarray}
\frac{1}{r}(r\tilde\phi)'' = 4\pi\l_B z_pN_{bp}/V + \kappa^2(\tilde \phi-\Delta\tilde\phi),   
\end{eqnarray}
where we have identified the internal (gel) charge density $-z_pN_{bp}/V = z_g c_g+c_s\exp(-\Delta\tilde\phi)-c_s\exp(\Delta\tilde\phi)$ from conditions (A5) and (A6), the internal concentration $c_{in} = c_g+c_s\exp(-\Delta\tilde\phi)+c_s\exp(\Delta\tilde\phi)$, and the internal inverse screening length $\kappa = \sqrt{4\pi\l_B c_{in}}$. 
The LPB equation can be solved with respect to the boundary condition that the electrical field on the sphere surface is fixed by 
$\tilde\phi'(R_p)=z_p l_B/R_p^2$ and it vanishes at the cell boundary $\tilde\phi'(R_c)=0$. 
The final solution is (see also previous works~\cite{cellmodel:denton,cellmodel:wittemann})
\begin{eqnarray}
\label{App_potential}
\tilde\phi(r)  &=& \Delta\tilde\phi+\tilde\phi_1(r) \\ \nonumber 
&=& \Delta\tilde\phi -z_pN_{bp}/(Vc_{in}) + A \frac{e^{-\kappa r}}{r} + B \frac{e^{\kappa r}}{r}
\end{eqnarray}
with constants
\begin{eqnarray}
A = \frac{z_p\l_B e^{\kappa R_p}}{1+\kappa R_p}\left[1-e^{-2\kappa (R_c-R_p)} \frac{(\kappa R_p -1)}{(\kappa R_p+1)}\frac{(\kappa R_c -1)}{(\kappa R_c+1)}\right]^{-1} \nonumber
\end{eqnarray}
and
\begin{eqnarray}
B = \frac{z_p\l_B}{1+\kappa R_p}\left[e^{\kappa(2R_c-R_p)}\frac{(\kappa R_c-1)}{(\kappa R_c +1)}-e^{\kappa R_p} \frac{(\kappa R_p -1)}{(\kappa R_p+1)}\right]^{-1} \nonumber
\end{eqnarray}
In the limit of large cell sizes, i.e., $R_c\rightarrow \infty$ (or $N_{bp}\rightarrow 0$), it follows that quickly 
$B\rightarrow 0$, and  $A\rightarrow z_p l_b \exp(\kappa R_p)/(1+\kappa R_p)$.   The LPB solution simplifies to  
\begin{eqnarray}
\tilde\phi(r)  = \Delta \tilde\phi - z_pN_{bp}/(Vc_{in})+\frac{z_p\l_B}{1+\kappa R_p} \frac{e^{-\kappa (r-R_p)}}{r}.
\label{potential_gel}
\end{eqnarray}
where we kept the important leading order term in $1/V$. An illustrative sketch of this potential distribution in the cell model 
is given  in Fig.~\ref{cellmodel}. $\Delta\phi$ is the average potential, while for small protein load the potential at the cell boundary 
$\phi(R_c)\simeq \Delta \tilde\phi -z_pN_{bp}/(Vc_{in})$ and at the particle surface $\phi(R_p)\simeq \Delta \tilde\phi -z_pN_{bp}/(Vc_{in}) +(z_p\l_B)/[R_p(1+\kappa R_p)]$. 
For not too high salt concentrations $c_s\lesssim c_g$, it follows that the Donnan potential $\Delta\phi\gtrsim 1$, 
and the internal salt concentration can be well represented by $c_{in} \simeq 2c_g$, as the coion 
concentration ($\propto \exp[-|\Delta\tilde\phi]|)$ in the gel becomes negligibly small. In the bulk solution in the dilute protein 
limit analogously to eq. (\ref{potential_gel}) we find
\begin{eqnarray}
\tilde\phi (r)  = \frac{z_p\l_B}{1+\kappa_b R_p} \frac{e^{-\kappa_b (r-R_p)}}{r},  
\end{eqnarray}
where $\kappa_b= \sqrt{8\pi\l_Bc_s}$ is the bulk inverse screening length. 

The transfer (Gibbs) free energy (or chemical potential) to bring a protein from bulk solution 
into the gel can then be calculated by the difference of work of charging the sphere against
the surface potential $\phi(R_p)$, $\beta\Delta G  = \int_0^{z_p} {\rm d}z_p[\phi_g(R_p)-\phi_b(R_p)] $  
in a cell in the gel (g) vs. bulk (b). We obtain the leading order contributions (in 
the low salt and low protein limit)  
\begin{eqnarray}
\label{App_transfer} 
\beta \Delta G_{el} &=& z_p\Delta\tilde\phi -\frac{z_p^2N_{bp}}{2c_gV}\\
 &-& \frac{z_p^2l_B}{2R_p}\left(\frac{\kappa_g R_p}{1+\kappa_g R_p}-\frac{\kappa_b R_p}{1+\kappa_b R_p}\right),\nonumber
\end{eqnarray}
where we defined $\kappa_g = \sqrt{8\pi\l_B c_g}$. We did not explicitly integrate over $z_p$ in the second term in (\ref{potential_gel}) as it is a constant background contribution not immediately involved in the charging process of one particle. 
It is noteworthy that we obtain the same functional 
form for $\Delta G_{el}$ if the  monomer charges are not assumed to be mobile, albeit with a $\sqrt2$ smaller 
internal inverse screening length  $\sqrt{4\pi\l_B c_g}$ and a factor of 2 in front of the second term. 
 
 By a one-to-one comparison of the leading order expression (\ref{App_transfer}) to the result from
 employing the full expression (\ref{App_potential}) we find that the error in $\Delta G_{el}$ is less than 
 $k_BT$ over the whole range of molar ratios and salt concentrations used in this work. By detailed
 inspection of the behavior of (\ref{potential_gel}) we observe that the reason of the surprising 
 accuracy is a fortuitous cancellation of errors of higher order terms at large $N_{bp}$. This fact 
 may shed some doubt on the general applicability of simplified eq.~(\ref{App_transfer}) but note
 that our parameters (protein valency, salt concentrations, monomer charge densities, etc.)
 are typical for a wide variety of experimental systems. However, in general the PB approach is 
 expected to break down for very high protein valencies and small proteins, when $|\tilde\phi_1|\gg1$, 
 and strong Coulomb correlations play a role.~\cite{hansen:mcdonald}

 The free energy (\ref{App_transfer}) above considers only ionic contributions to solvation of
 a fixed lattice of spheres, i.e., it neglects the electrostatic contributions from the interaction between 
 the proteins, i.e., the energy penalty of overlapping double layers. However, in the fluid-like hydrogel protein matrix 
 it is reasonable to assume that proteins can wiggle or move around and are not rigidly fixed to lattice 
 positions. Due to such fluctuations 
  the average surface potential will actually be higher than given in (\ref{App_potential}).
 To estimate the interaction contribution we look at the expansion of the excess chemical potential 
 in terms of virial coefficients, i.e, in first order 
 \begin{eqnarray}
 \beta\mu = 2B_2/V_c, 
 \end{eqnarray}
where $B_2 = -\frac{1}{2}\int {\rm d}^3r\;[\exp(-\beta W(r)]-1)$ and 
\begin{eqnarray}
W(r) = W_{HS}+\frac{z_p^2\l_B}{1+\kappa R_p} \frac{e^{-\kappa (r-R_p)}}{r}
\end{eqnarray}
is the protein-protein interaction potential, split up into hard-sphere and the Debye-H\"uckel contribution 
according to  (\ref{App_potential}). Using condition (\ref{norm2}), linearizing the exponent in the defining equation for
$B_2$,  and splitting the chemical potential into hard-sphere and electrostatic contributions we find
 \begin{eqnarray}
 \beta\mu = \beta \mu_{HS} + \frac{z_p^2N_{bp}}{2c_gV}. 
 \end{eqnarray}
Thus in leading order the electrostatic protein-protein interaction contribution exactly cancels the second term in 
(\ref{App_transfer}) and we end up with the final result for the electrostatic transfer free energy
\begin{eqnarray}
\beta \Delta G_{el} = z_p\Delta\tilde\phi - \frac{z_p^2l_B}{2R_p}\left(\frac{\kappa_g R_p}{1+\kappa_g R_p}-\frac{\kappa_b R_p}{1+\kappa_b R_p}\right)
\label{App_transfer_final} 
\end{eqnarray}
It is interesting to note on the simplicity of eq.~(\ref{App_transfer_final}) which describes naively 
the transfer of charge $z_p$ into the average potential $\tilde\Delta\phi$ and, in the second term, 
the difference in Born solvation free energies in a homogeneous medium with 
salt concentrations $c_s$ and $c_g$. Thus, the inhomogeneities introduced by the cell 
model assumption as depicted in Fig.~5 cancel out (in linearized theory) if particle 
fluctuations are allowed, and the naive form (\ref{App_transfer_final}) holds.
    
The estimation of ionic osmotic contribution to pressure in the presence of proteins seems less simple.
Microconfiguration of proteins induce inhomogeneities and  the pressure is not anymore given by the 
simple expression for $\pi_{\rm}$ in eq.~(\ref{ionpressure}). The ion contribution to the osmotic pressure 
is usually estimated from the ionic concentration at the cell surface (in the cell model) where the electrostatic pressure vanishes,~\cite{cellmodel:marcus, cellmodel:pincus, cellmodel:denton, biesheuvel:pre,vos:langmuir,cellmodel:wittemann} 
\begin{eqnarray}
\pi_{ion}^p \simeq c_s\exp(\tilde\phi(R_c)) +c_s\exp(-\tilde\phi(R_c)) -2c_s.
\end{eqnarray}
Expanding to 1st order in $\phi_1$,  we obtain
\begin{eqnarray}
\pi_{ion}^p \simeq \pi_{ion}  +2c_s\tilde\phi_1(R_c)\sinh(\Delta\tilde\phi), 
\end{eqnarray}
which recovers $\pi_{ion}^p \simeq \pi_{ion}$ in the limit for vanishing protein
concentrations and $\Delta\tilde\phi(y)$ is the protein-corrected Donnan potential (\ref{donnan_mod}) with $y = (z_g c_g + z_pN_{bp}/V)/(2c_s)$. Note that this expression does not consider fluctuations in protein positions which is likely
to be an important effect to consider in future studies. 
 
\section{The standard Langmuir model in the canonical ensemble}

Consider a finite region in space with $N$ identical and independent binding sites 
available. We denote the number of bound proteins by $N_{bp}$ and 
define the fraction of bound particles by $\Theta=N_{bp}/N$. 
The number of available binding states is then~\cite{langmuir,langmuir2}
\begin{eqnarray}
W = \frac{\zeta^{N_{bp}} N!}{N_{bp}!(N-N_{bp})!}, 
\end{eqnarray}
from the combinatorial possibilities of distributing $N_{bp}$ indistinguishable particles on $N$ sites, and
$\zeta$ is the partition sum of a single particle in the bound state.  The Boltzmann entropy is defined by 
\begin{eqnarray}
\frac{S}{k_B}=\ln W 
\end{eqnarray}
leading (within a constant) to the entropy per binding site 
\begin{eqnarray}
\frac{S}{N\,k_B} = -\Theta\ln\Theta-(1-\Theta)\ln(1-\Theta)+\Theta \ln(v_0/\Lambda^3), 
\end{eqnarray}
where we defined $\zeta^{N_{bp}}=(v_0/\Lambda^3)^{N_{bp}}$ in terms of an effective configurational volume $v_0$
divided by the cubed thermal (de Broglie) wavelength $\Lambda^3$.  'Effective' means that also restrictions on 
vibrational and orientational degrees of freedom are adsorbed in the number $v_0$, not purely translational 
effects if it all. The (canonical) Helmholtz free energy of the system is
\begin{eqnarray}
\beta F = \beta F_{id}-\frac{S }{k_B}+\beta N_{bp}\Delta G, 
\end{eqnarray}
where we introduced the canonical ideal gas free energy $\beta F_{id}=(n-N_{bp})[\ln((n-N_{bp})\Lambda^3/V)-1]$ with 
total particle number $n$. Then $(n-N_{bp})/V$ is the density of unbound particles in a total volume $V$ (assumed to be much larger than the binding region),  and the adsorption 
free energy $\Delta G$  associated with the binding of one protein to one Langmuir site.  
The free energy $\tilde f=\beta F/N$ per binding site is then
\begin{eqnarray}
\tilde f &=& \frac{1}{N}(n-N_{bp})[\ln((n-N_{bp})\Lambda^3/V)-1] \\ \nonumber &+&\Theta\ln\Theta+(1-\Theta)\ln(1-\Theta)\\&-&\Theta \ln(v_0/\Lambda^3)+\beta \Theta\Delta G. \nonumber
\end{eqnarray}
The minimization of the free energy  with respect to the number of bound 
protein $\partial\tilde f/\partial N_{bp} =0$ yields then the final relations for the fraction of bound
particles in dependence of the (unbound) particle concentration $c_p=(n-N_{bp})/V$.
We obtain the final result
\begin{eqnarray}
K= \exp(-\beta\Delta G)v_0=  \frac{\Theta}{(1-\Theta)c_p}.
\end{eqnarray}
The standard volume $v_0$ depends on the exact nature of the bound state and is typically not known. 
Thus the prediction of absolute binding free energies is difficult. In literature often the
standard volume l/mol$\simeq 1.6$~nm$^3$ is employed. This is not unreasonable if
one assumes that still rotational and vibrational modes in the bound state take place
on molecular, i.e, nanometer scales.

\newpage

\begin{figure}[h]
\centerline{\includegraphics[width=10cm,angle=0]{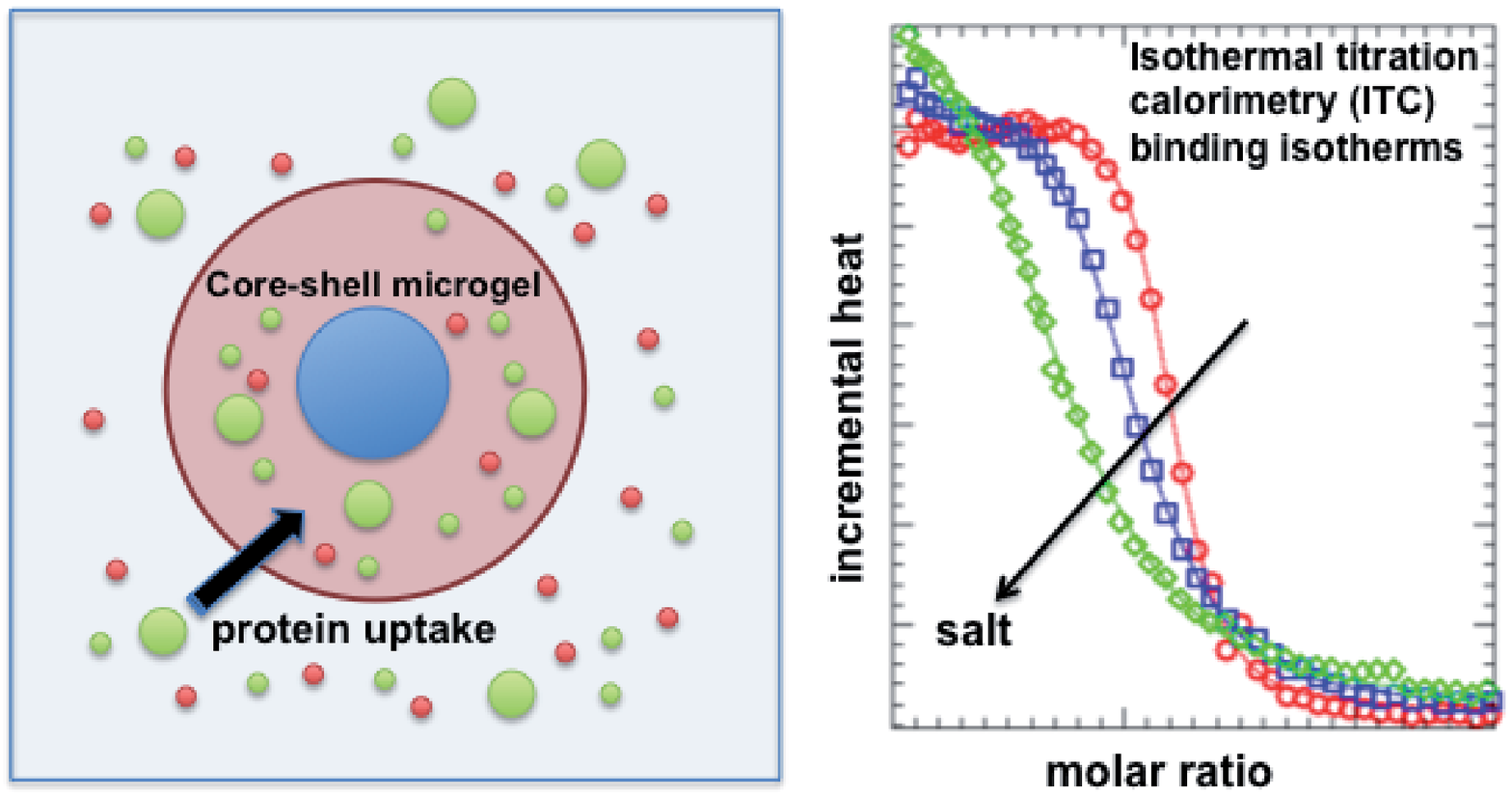}}
\caption{TOC figure}
\end{figure}
\end{document}